\documentclass[10pt,conference]{IEEEtran}
\IEEEoverridecommandlockouts
\usepackage{cite}
\usepackage{amsmath,amssymb,amsfonts}
\usepackage{algorithmic}
\usepackage{graphicx}
\usepackage{textcomp}
\usepackage{xcolor}
\usepackage{tabularx}
\usepackage{booktabs}
\usepackage{multirow}
\usepackage{adjustbox}
\usepackage{alltt}
\usepackage{listings}

\usepackage{pdflscape}
\usepackage{array, multirow}

\usepackage{subcaption}
\usepackage{fancyhdr}
\usepackage{longtable}
\usepackage{array}

\usepackage{balance}

\def\BibTeX{{\rm B\kern-.05em{\sc i\kern-.025em b}\kern-.08em
    T\kern-.1667em\lower.7ex\hbox{E}\kern-.125emX}}

\begin{document}

\title{Checkstyle+: Reducing Technical Debt Through The Use of Linters with LLMs\\
}

\author{\IEEEauthorblockN{Ella Dodor}
\IEEEauthorblockA{\textit{Department of Informatics} \\
\textit{University of California, Irvine}\\
Irvine, California, US \\
edodor@uci.edu}
\and
\IEEEauthorblockN{Cristina V. Lopes}
\IEEEauthorblockA{\textit{Department of Informatics} \\
\textit{University of California, Irvine}\\
Irvine, California, US \\
lopes@uci.edu}
}

\maketitle

\begin{abstract}
Good code style improves program readability, maintainability, and collaboration, and is an integral component of software quality. Developers, however, often cut corners when following style rules, leading to the wide adoption of tools such as linters in professional software development projects. Traditional linters like Checkstyle operate using rigid, rule-based mechanisms that effectively detect many surface-level violations. However, in most programming languages, there is a subset of style rules that require a more nuanced understanding of code, and fall outside the scope of such static analysis. In this paper, we propose Checkstyle+, a hybrid approach that augments Checkstyle with large language model (LLM) capabilities, to identify style violations that elude the conventional rule-based analysis. Checkstyle+ is evaluated on a sample of 380 Java code files, drawn from a broader dataset of 30,800 real-world Java programs sourced from accepted Codeforces~\cite{Codeforces} submissions. The results show that Checkstyle+ achieves superior performance over standard Checkstyle in detecting violations of the semantically nuanced rules.
\end{abstract}


\section{Introduction}

Code style is not just a matter of aesthetics, it plays a crucial role in program readability~\cite{Oliveira2023}, maintainability~\cite{Coleman2018Beauty}, and collaboration~\cite{Zou2019}; furthermore inconsistent code style practices can contribute to the accumulation of technical debt (TD). Technical debt is a metaphor for the long-term costs associated with taking shortcuts or making suboptimal decisions in software development~\cite{Kruchten2013TechDebt}. Inconsistent code style can introduce friction during development, increase onboarding time for new contributors~\cite{Haque2025GameChanger}, and exacerbate long-term maintenance costs. From a technical debt perspective, code style issues can be classified as a form of ``code TD" or ``architectural TD"~\cite{Li2015TechDebtMapping}. Li et al.~\cite{Li2015TechDebtMapping} classify ``code TD" as debt that arises from suboptimal implementation choices at the code level, including poor code style and over-complex code. As for ``architectural TD", their study defines it as the consequences of architectural decisions that trade off certain internal quality attributes, such as maintainability, in favor of other short-term goals. They also take a note of structural quality issues, which was not classified as a type of TD but nonetheless can be merged into ``architectural TD"~\cite{Li2015TechDebtMapping}. While technical debt can be hard to measure, its impact on developers is impossible to ignore. At the project level, TD is linked to increased development and maintenance costs, delayed feature delivery, and reduced software quality~\cite{Besker2019ProductivityDebt,paudel2024measuringimpacttechnicaldebt,Ramasubbu2016TechDebtReliability}. Empirical studies consistently show that developers must spend a significant portion of their time mitigating the effects of TD instead of building new functionality~\cite{Besker2019ProductivityDebt}. Given this, many efforts have been made towards managing and reducing technical debt: in this effort Li et al. identified 8 activities and 29 tools for technical debt management, one of which was Checkstyle~\cite{checkstyle}, a popular Java linter, which we augmented with LLM capabilities in this paper.

\begin{figure}
    \centering
    \includegraphics[width=\linewidth]{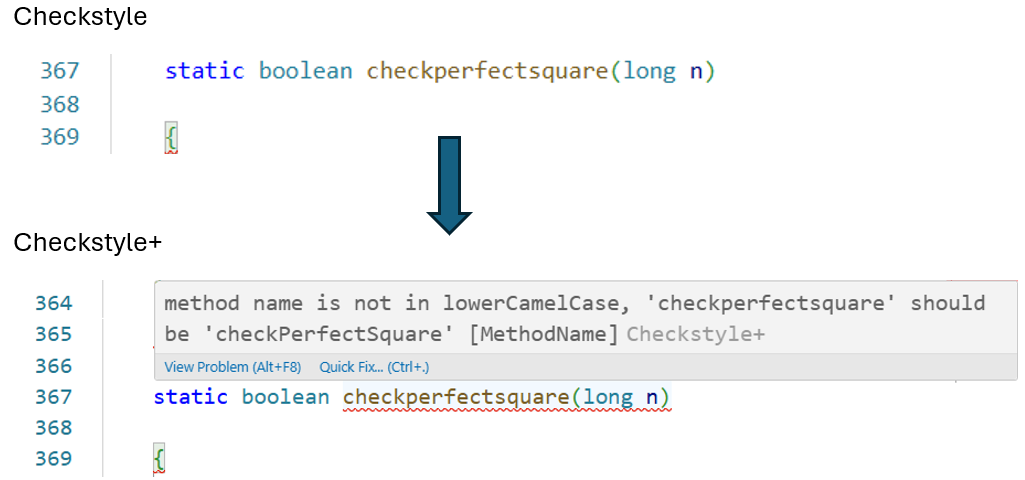}
    \caption{Example of Checkstyle's False Negatives Caught by Checkstyle+}
    \label{fig:exampleFalseNegativeIntro}
\end{figure}

Traditional linters such as Checkstyle~\cite{checkstyle} have been widely adopted to ensure adherence to established style guidelines, and help reduce stylistic code defects that could lead to technical debt. While effective, traditional linters can only provide rule-based checks, and are limited when it comes to identifying subtle deviations that require using context, recognizing patterns and a deeper understanding of language~\cite{Dodor2024LLMs}. For example, the Google Java Style Guide~\cite{JavaGuide} recommends that method names follow lowerCamelCase formatting, where the first word begins with a lowercase letter, but each subsequent word begins with a capital letter. Checkstyle fails to flag violations like ``checkperfectsquare" (see Figure~\ref{fig:exampleFalseNegativeIntro}), which appears syntactically valid due to being entirely lowercase, but semantically violates the guideline by not capitalizing the internal word boundaries. Because Checkstyle, and linters, in general, rely solely on formal language rules and lack natural language understanding, they has no mechanism for determining where one word ends and another begins. As seen in the previous example, many of these nuanced deviations, particularly in identifier naming and documentation quality, are difficult to catch through regular expressions or formal rules; yet they are among the most influential factors affecting code readability and maintainability~\cite{Binkley2013IdentifierStyle,Carpio2017NamingPatterns,Oliveira2023,Coleman2018Beauty}.

Large Language Models (LLMs) are capable of understanding natural language, identifying stylistic patterns, and reasoning about code beyond syntax. Prior work~\cite{Dodor2024LLMs} on the use of LLMs for code style enforcement shows that LLMs are not replacements for linters, because they fail at detecting violations of many simple stylistic rules, such as indentation and white space. However, they present key advantages in stylistic rules involving natural language understanding. Given this, we present Checkstyle+, an extension of Checkstyle with LLM capabilities that are invoked only for specific rules: those requiring a deeper understanding of language. To assess and understand the impact of this augmentation, Checkstyle+ is compared to Checkstyle using a dataset of real-world developer code. In this work, we answer the following research questions:

\begin{enumerate}
    \item Can Large Language Models (LLMs) improve coverage of style violations related to naming and documentation, beyond what traditional linters detect?
    \item What are the trade-offs of LLM-based style enforcement?
    \item To what extent does model non-determinism impact the reliability of its style recommendation?
\end{enumerate}

The remainder of this paper is organized as follows. Section~\ref{sec:background} provides background on technical debt and code style enforcement, followed by related work in this domain. Section~\ref{sec:checkstyleplus} outlines the design choices of Checkstyle+, including targeted rules, LLM integration, configurable properties, and other decisions. Section~\ref{sec:evaluation} details the methodology for comparing the performance of Checkstyle+ with Checkstyle, and finally, Section~\ref{sec:res} presents the results, followed by a Discussion (Section~\ref{sec:discussion}), Threats to Validity (Section~\ref{sec:ttv}) and the Conclusion (Section~\ref{sec:conclusion}).

\section{Background}
\label{sec:background}

Code style is a fundamental aspect of software development; it is crucial for readability~\cite{Oliveira2023}, maintainability~\cite{Coleman2018Beauty}, and collaboration~\cite{checkstyle} across teams. Consistent code style practices make code easier to understand, reduce the likelihood of errors, and facilitate long-term project evolution~\cite{findingArt, Oliveira_2023}. Over the years, a variety of tools, commonly referred to as linters, have been developed to automatically detect and report deviations from established style guidelines. Tools like Checkstyle~\cite{checkstyle}, Flake8~\cite{flake8}, and ESLint~\cite{ESLint} apply rule-based checks that are efficient and reliable for syntactic violations, ensuring that basic formatting and structural conventions are followed. Style linters have been identified as tools that help manage technical debt~\cite{Li2015TechDebtMapping}, however, they present certain limitations that have pushed researchers to look into Machine Learning (ML), approaches to style enforcement~\cite{Style-Analyzer, naturalize}.

\subsection{Code and Architectural Technical Debt}

Technical debt (TD) can manifest at multiple levels or phases of the software development process, but in this paper we focus mainly on code TD. Code TD typically arises from poor coding practices, including overly complex logic, inadequate documentation, or inconsistent code style, which collectively reduce code readability, increase maintenance effort, and affect future development~\cite{Li2015TechDebtMapping}. Lenarduzzi et al.~\cite{Lenarduzzi2021TechDebtPrioritization} argue that code and architectural are among the most investigated types of TD because of their direct and measurable impact on development activities.

Many works have focused on managing technical debt through structured processes, practices, and automation. Recent studies highlight that managing TD, particularly at the code level—increasingly involves tool, supported strategies that assist with detection, prioritization, and remediation. For example, Biazotto et al.~\cite{Biazotto2024TechDebtAutomation} provide an overview of automated techniques used across the TD lifecycle, underscoring how automation has become a key enabler for scalable and systematic TD management.

More recent research has begun to investigate how large language models (LLMs) can support in the management of technical debt. For instance, works like Augmented Code Engineering (ACE)~\cite{Tornhill2025ACE} offers LLM generated refactorings with human validation within developers’ IDE workflows, targeting code smells. Other works like those of Sheikhaei et al.~\cite{Sheikhaei2025LLMTD} investigate how large language models can support the automated repayment of self-admitted technical debt (SATD). This shift highlights a growing trend toward active automation of debt repayment, using the capabilities of LLMs to assist developers in maintaining and improving code quality.

\subsection{Code Style Enforcement Tools}

Code style enforcement has long relied on linters, to automatically detect deviations from established conventions. Tools such as Flake8~\cite{flake8} for Python, Checkstyle~\cite{checkstyle} for Java, and ESLint~\cite{ESLint} are widely adopted in both industry and education. These linters operate by applying a predetermined set of rules to source code, flagging issues like improper indentation, incorrect naming, or missing whitespace. By automating these checks, linters reduce the burden on human reviewers and help maintain consistency across large projects. Their efficiency and deterministic nature make them indispensable for enforcing style rules.

\begin{figure}
    \centering
    \includegraphics[width=\linewidth]{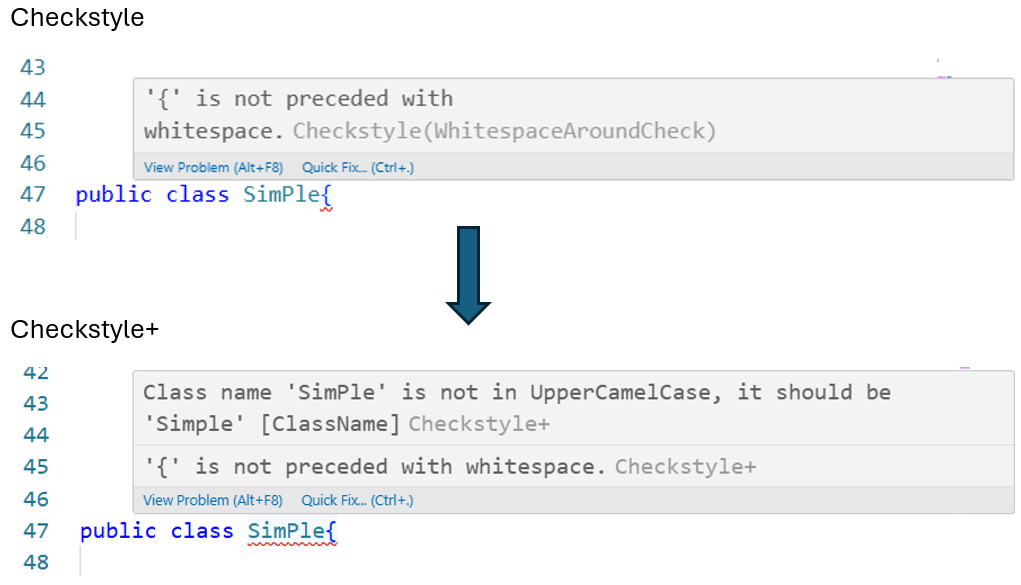}
    \caption{Incorrect Capitalization: Checkstyle vs. Checkstyle+}
    \label{fig:capitalizationExampleBG}
\end{figure}

\begin{figure}
    \centering
    \includegraphics[width=1\linewidth]{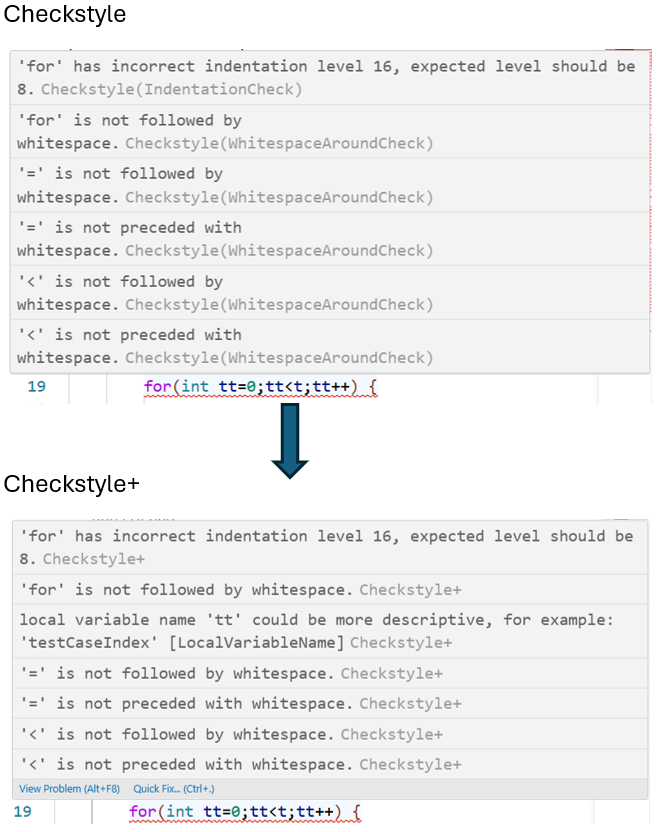}
    \caption{Identifier Descriptiveness: Checkstyle vs. Checkstyle+ }
    \label{fig:descriptivenamebg}
\end{figure}

However, the rule-based foundation of traditional linters introduces significant limitations. Because they operate on explicitly codified rules, linters can only identify violations that can be expressed in patterns. While this works well for enforcing conventions like maximum line length or brace placement, it fails when style guidelines require reasoning about broader patterns or context. For instance, Checkstyle may verify that a class name begins with a capital letter but cannot determine whether the rest of the identifier is correctly capitalized (see Figure~\ref{fig:capitalizationExampleBG}), or whether a variable name is meaningful and descriptive within its context (see Figure~\ref{fig:descriptivenamebg}). 

In recent years, research has explored alternatives that move beyond strict rule-based enforcement by leveraging data-driven and machine learning methods. Tools such as NATURALIZE~\cite{naturalize} and STYLE-ANALYZER~\cite{Style-Analyzer} learn naming and formatting patterns from large code corpora to provide style suggestions tailored to a project. Others, like GRADESTYLE~\cite{GRADESTYLE}, integrate automated feedback directly into developer workflows to encourage gradual style improvements over time. While these approaches reduce reliance on manually written rules, they, too, face challenges: statistical methods may miss semantic cues, and workflow-integrated tools often introduce delays or lack qualitative feedback. Together, these developments illustrate both the progress and the persistent gaps in code style enforcement, motivating the exploration of more context-aware techniques such as Large Language Models (LLMs).

\subsection{LLMs for Code Style Analysis}

In recent years, LLMs have been increasingly applied across a wide spectrum of software engineering tasks. Zhang et al.’s~\cite{zhang2024surveylargelanguagemodels} recent survey reports 947 studies by mid-2024 that explore LLM use in 112 code-related tasks spanning five SE phases: requirements and design, development, testing, maintenance, and management. This increase in LLM usage is starting to also be observed when it comes to static analysis tool. For example, Li et al. integrates ChatGPT with UBITect to resolve ambiguous bug cases caused by resource limitations~\cite{Li2023LLMStaticAnalysis}. Other studies also follow this growing trend, in slightly different manners, for example, Woodrow et al.~\cite{Woodrow2024ElegantCoding} show how LLMs can provide real-time, context-aware style feedback to computer science students. Another study compared the capabilities of both linters and LLMs, for enforcing code style in Python and Java code. The results demonstrated that while traditional linters generally outperformed LLMs in terms of raw accuracy and reliability, they showed distinct strengths in areas where style rules required contextual or semantic reasoning. Linters excelled in formatting-related checks and rule based programmable detections~\cite{Dodor2024LLMs}.

\section{Checkstyle+}
\label{sec:checkstyleplus}
This section describes the core architectural and implementation decisions behind building Checkstyle+: the configuration mechanisms, the prompt–response workflow, and how the tool handles and classifies recommendations. This design emphasizes modularity, reproducibility, compatibility and consistency with existing Checkstyle infrastructure.
The implementation of Checkstyle+, along with configuration files and scripts used in our evaluation, is publicly available at https://github.com/ellacodee/CheckstylePlus.

\subsection{Targeted Style Guidelines}

This study focuses on code style guidelines from the Google Java style guide~\cite{JavaGuide} that require deeper semantic and contextual understanding, in which LLMs have demonstrated promise in the past~\cite{Dodor2024LLMs}. In particular, we target naming rules related to class names, method names, constants, non-constant fields, parameters, local variables, and documentation rules related to Java doc formatting and summary fragments. These aspects were selected because their enforcement often depends on additional understanding of the structure, intent, and meaning of identifiers or documentation, rather than only their surface form. Table~\ref{tab:guidelines} presents a total of 9 targeted guidelines, structured into two sections, documentation (javadoc summary fragment and implementation comment formatting),  and naming conventions (class, method, constants and non-constants, parameter, local variable and type variable names). All other remaining guidelines are handled by the default Checkstyle checks (i.e., line length, spacing, imports, general formatting)

\begin{table*}[t]
\centering
\caption{Summary of targeted Google Java Style Guide rules.}
\renewcommand{\arraystretch}{1.2}
\setlength{\tabcolsep}{6pt}
\begin{tabular}{|p{3.1cm}|p{3.5cm}|p{10cm}|}
\hline
\textbf{Category} & \textbf{Subcategory} & \textbf{Guideline} \\
\hline

\multirow{1}{2.5cm}{1-Documentation}
& \multirow{1}{4.2cm}{1.1-Javadocs}
& 1.1.1-Each Javadoc block begins with a brief summary fragment. This fragment is very important: it is the only part of the text that appears in certain contexts such as class and method indexes. The fragment is capitalized and punctuated as if it were a complete sentence. \\
\cline{3-3}
& & 1.1.2-Implementation comments defining purpose or behavior are written as Javadoc. \\
\hline

\multirow{1}{3.1cm}{2-Naming Conventions}
& \multirow{1}{4.2cm}{2.1-Class Names}
& 2.1.1-Class names are written in UpperCamelCase. \\
\cline{2-3}
& \multirow{1}{4.2cm}{2.2-Method Names}
& 2.2.1- Method names follow specific camelCase rules. \\
\cline{2-3}
& \multirow{2}{4.2cm}{2.3-Constant Names}
& 2.3.1-Constant names use UPPER\_SNAKE\_CASE. \\
\cline{3-3}
& & 2.3.2-Non-constant field names use lowerCamelCase. \\
\cline{2-3}
& \multirow{1}{4.2cm}{2.4-Parameter Names}
& 2.4.1-Parameter names are written in lowerCamelCase. \\
\cline{2-3}
& \multirow{1}{4.2cm}{2.5-Local Variables}
& 2.5.1-Local variable names follow specific camelCase rules. \\
\cline{2-3}
& \multirow{1}{4.2cm}{2.6-Type Variable Names}
& 2.6.1-Each type variable is named in one of two styles: \\
& & \hspace{8.5mm}— A single capital letter optionally followed by a numeral (e.g., E, T, X, T2). \\
& & \hspace{8.5mm}— A name in class-name form followed by T (e.g., RequestT, FooBarT). \\
\hline

\end{tabular}
\label{tab:guidelines}
\end{table*}

\subsection{Model and Prompt}

Checkstyle+ relies on an LLM for recommendations concerning the 9 guidelines presented in Table~\ref{tab:guidelines}. Checkstyle+ uses gemini-2.5-pro as its default model due to its high precision (0.99), strong recall (0.81), and the best overall F1 score (0.89), among the evaluated models. While gemini-2.5-pro is the default model, Checkstyle+ is designed to be flexible and can be configured to work with other models. 

An iterative refinement process was used to select the best prompt for querying the model used in Checkstyle+. The prompt was applied to a representative sample of 10 Java code files and analyzed against a baseline of existing violations (56 total) to identify true positives, false positives and false negatives. The prompt was also adjusted to improve clarity, and reduce ambiguity in the model's response. This cycle was repeated until the model consistently produced structured and accurate outputs that matched the expected ground truth. The resulting prompt was structured to make the model’s outputs traceable and unambiguous. Each violation in the response was required to include both the relevant guideline section number and the corresponding line number, which improved the model’s ability to precisely locate and label problematic tokens. To support an accurate interpretation, the prompt included a combination of clear task instructions, explicit output formatting rules, carefully selected examples of targeted violations, a description of the 9 selected guidelines, followed by the code to be analyzed. Figure~\ref{fig:prompt-preview} shows a preview of the structure of the prompt, and the full prompt can be found in Appendix~\ref{sec:prompt}. In addition to structuring the prompt, we introduced a simple classification system to make the model’s outputs more expressive and reduce false positives. Clear-cut violations of the targeted guidelines were labeled as errors, while stylistic or readability issues not explicitly stated in the guidelines were labeled as warnings. This distinction allowed the model to report non-critical findings without inflating the number of false positives and provided a clearer separation between strict violations and softer recommendations.

\lstset{
  basicstyle=\ttfamily\scriptsize,
  breaklines=true,
  breakatwhitespace=true,
  columns=fullflexible,
  frame=single,
  numbers=none
}

\begin{figure}[t]
\centering
\begin{minipage}{0.95\columnwidth}
\begin{lstlisting}
TASK:
Analyze the code for violations of only these sections 
guidelines: 1.1.1, 1.1.2, 2.1.1...

OUTPUT RULES:
Report each violation exactly as:
[Error](line number) (violation section number) ...

CAMEL CASE NAMING GUIDE:
To convert a name into prper camel case...

JAVA STYLE GUIDELINES:
1- Documentation...
2- Naming Convention...

JAVA CODE:
\end{lstlisting}
\end{minipage}
\caption{Excerpt from the LLM prompt used for code style analysis. See Appendix~\ref{sec:prompt} for the full prompt.}
\label{fig:prompt-preview}
\end{figure}

\subsection{Architecture}

Checkstyle+ is implemented as a custom Checkstyle module by extending the AbstractCheck base class. This approach leverages the existing static‑analysis pipeline provided by Checkstyle, allowing the new functionality to be injected without modifying the tool’s core. Whenever Checkstyle traverses the abstract syntax tree (AST) of a Java file, the custom module is invoked on the file and can log the targeted violations just like native checks. All other structural checks like line length, spacing issues or indentation are still handled by Checkstyle's default checks; Checkstyle+ only focuses on 2 documentation related guidelines, and 7 naming related guidelines (Table~\ref{tab:guidelines}), that require context and a deeper language understanding.

\subsection{Configurable Properties and User Control}

Checkstyle includes a configuration file that allows users to customize which rules to include in their checks as well as other properties. Likewise, Checkstyle+ exposes an extra set of configurable properties that allow users to tailor the tool’s behavior entirely through the checkstyle.xml configuration file. The key properties include:

\begin{itemize}
    \item \textbf{apiKey and endpoint}: These specify the authentication token and endpoint for the model selected by the user. By default, the tool is set up for gemini-2.5-Pro's API.

    \item \textbf{model (optional)}: For some APIs (e.g., Gemini), the model name is already embedded in the endpoint URL and does not need to be provided separately. For others (e.g., OpenAI), the model must be specified explicitly.
    
    \item \textbf{temperature}: Controls the level of randomness in the model’s output. By default the temperature is set to 1.

    \item \textbf{maxTokens}: Sets the upper limit on the number of tokens returned by the model in a single response. Inherently this is set to the default output cap of the model, however this could be changed to reduce cost.

    \item \textbf{thinkingTokens}: Configures the number of tokens allocated to the model’s internal reasoning before generating a final response. By default there is no thinking budget set, meaning the model automatically allocated an internal number of reasoning tokens, however this can be changed to reduce cost.
    
    \item \textbf{showWarnings}: When set to false, Checkstyle+ filters out non-critical style recommendations ``warnings" returned by the model, allowing users to focus only on strict guideline violations, and reduce unnecessary noise.

    \item \textbf{enabled}: This boolean flag allows toggling the entire check on/off at runtime, which is especially useful for comparative experiments, and allows the user to default to just using Checkstyle without the LLM capabilities due to cost.
    
\end{itemize}

\subsection{Caching}

When a code file is first sent to Checkstyle+ for analysis, the model’s response is hashed and placed in a cache. If the same file is analyzed again, the tool reuses the saved response instead of calling the model again. This makes the results consistent across runs, reduces latency, and lowers API costs.

\subsection{Response Handling, Classification, and Logging}

\begin{figure*}
    \centering
    \includegraphics[width=1\linewidth]{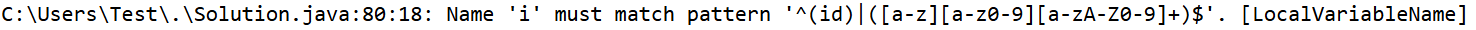}
    \caption{Checkstyle's Standard Output}
    \label{fig:checkstyleoutput}
\end{figure*}

It is important that the rules relying on LLMs be seamlessly integrated with how Checkstyle reports all other rules. To achieve this, the model’s output is processed through the handleLLMResponse() method, which reads the response line by line and identifies severity levels based on prefixes such as [ERROR] and [WARNING]. Depending on the `showWarnings' setting, warnings can either be displayed in Checkstyle+'s output, or suppressed entirely. The parsed recommendations are then logged through \texttt{log(line, column, message)}, to match Checkstyle's output format. Checkstyle reports each issue in a standardized structure that includes a severity level, the file path, line and column number, a short message describing the violation, and a descriptive tag that contains the name of the affected rule (i.e., MethodName, TypeName, or MemberName), see Figure~\ref{fig:checkstyleoutput}.

To align with Checkstyle’s descriptive tags, each LLM recommendation is mapped to the appropriate Checkstyle descriptor corresponding to the targeted guidelines: [SummaryJavadoc], [MethodName], [ParameterName], [LocalVariableName], [MemberName], and [ClassTypeParameterName]. Because Checkstyle lacks a dedicated descriptor for class name violations (uses [TypeName], a new descriptor, [ClassName], is introduced. Similarly, for incorrectly formatted implementation comments describing class or member behavior, [JavadocRequired] is added. 

\section{Evaluation of Checkstyle+}
\label{sec:evaluation}
\subsection{Dataset Selection}
To compare Checkstyle+ with Checkstyle, a dataset reflecting real-world coding style practices is selected. The tools are compared using a Kaggle dataset~\cite{codeforces_kaggle_accepted_submissions} consisting of 107,000 accepted Python, Java and C++ source code submissions. These were drawn from 3,000 different Codeforces~\cite{Codeforces} problems, primarily from contests held between 2020 and 2023. 28.85\% of the total code submissions were in Java, which resulted in a dataset of 30,867 Java code files. This ensures that the evaluation is grounded in realistic coding behaviors, making the findings more representative of how style enforcement tools would perform on developer-written code.

From the 30,867 Java submissions, we determined a representative sample size using a 95\% confidence level and a 5\% margin of error, resulting in a final sample of 380 randomly selected code files for our study. 

\subsection{Baseline} To ensure a fair and replicable comparison between Checkstyle+ and Checkstyle, we first established a ground truth dataset of verified style violations relating to 9 targeted guidelines: javadoc summary fragments (1.1.1), nuanced implementation comment formatting (1.1.2), class names (2.1.1), method names (2.2.1), constant names (2.3.1), non-constant field names (2.3.2), parameter names (2.4.1), local variable names (2.5.1), type variables (2.6.1) (see Table~\ref{tab:guidelines}). Checkstyle’s default Google Java~\cite{JavaGuide} style rule configurations were applied to the 380 sampled Java files to identify candidate violations. These outputs were then manually reviewed by a human evaluator to eliminate false positives and to annotate cases where Checkstyle failed to detect violations. The resulting set of validated labels served as the reference standard against which both tools were evaluated.

\subsection{Model Selection}

To determine the most suitable model for Checkstyle+, we conducted a comparative evaluation of several state-of-the-art LLMs, including Gemini 2.5 Flash~\cite{google_gemini_thinking}, Gemini 2.5 Pro~\cite{google_gemini_thinking}, o3-2025-04-16~\cite{OpenAIProducts}, and GPT-5~\cite{OpenAIProducts}. We specifically targeted reasoning models because of their improved accuracy, reasoning and multi-step planning abilities. Each model was queried 5 times with the same set of 10 representative Java code files, containing 56 style violations, and their outputs were manually analyzed. All models achieved very high precision (\~0.99), indicating a low rate of false positives; the primary differentiator was recall. Gemini 2.5 Pro obtained the highest recall (0.807) and the best F-measure (0.892), with the lowest false-positive rate tied (0.2), outperforming GPT-5 (F1 0.870), Gemini 2.5 Flash (F1 0.831), and o3-2025-04-16 (F1 0.746). Based on this balance of precision and recall, we selected Gemini 2.5 Pro for the study.

\subsubsection{Temperature}

To evaluate the impact of temperature on model performance, we conducted a comparative study of 5 temperature settings (0 to 1 in 0.25 increments) using gemini-2.5-pro. Each setting was tested over 5 iterations of 10 representative Java files, containing 56 style issues. Precision remained near-perfect across all settings, ranging from 0.995 to 1, indicating that increased sampling diversity did not introduce false positives. The most notable variation occurred in recall, which increased from 0.667 at temperature = 0 to 0.879 at temperature = 1, despite a slight dip at temperature = 0.25. Correspondingly, false negatives fell from an average of 18.6 to 6.8. F-measure followed a similar trend, improving from 0.8 at temperature = 0, to 0.94 at temperature = 1. Based on this balance of coverage and precision, we selected temperature = 1 for the study.

\subsubsection{Max Output Tokens}

We kept the default output cap for Gemini 2.5 Pro and verified that no responses were truncated, because truncation can cause false negatives (missed violations) if the model’s explanation is cut off before listing all findings. Since max tokens limits only the output length, the priority was simply to prevent length-based cutoffs that would bias recall. In the runs, no outputs ended due to length, so increasing the cap further was unnecessary. According to Google~\cite{google_gemini_thinking}, Gemini 2.5 Pro’s maximum output tokens default is 65,535, which comfortably exceeded observed needs.

\subsubsection{Thinking Tokens}

We used the default thinking tokens setting provided by the model, which allocates a fixed number of tokens for internal reasoning. This choice avoids introducing additional variability and ensures consistency with the model’s standard reasoning behavior.

\subsection{Evaluation Setup} 
Checkstyle and Checkstyle+ were both executed on the same 380 code samples, focusing on 2 documentation guidelines (1.1.1 and 1.1.2), and 7 naming guidelines (2.1.1, 2.2.1, 2.3.1, 2.3.2, 2.4.1, 2.5.1, and 2.6.1), shown in Table~\ref{tab:guidelines}. Outputs were compared to the established ground truth using human evaluators. A match between a predicted and true violation was treated as a true positive; a detected violation absent from the ground truth was marked as a false positive; and any missed ground-truth violation was recorded as a false negative. Using these counts, we computed precision, recall, and F1 scores for each tool to determine overall tool performance. In addition to quantitative evaluation, we performed a qualitative assessment of both tools' outputs, focusing on areas of differences, clarity, and instructional value.

We also measured both the cost and latency associated with running each tool, as well as the stability of LLM outputs across runs. Since Checkstyle is a static, deterministic tool with no associated usage fees, only latency was measured for this baseline. For Checkstyle+, we measured both cost and latency. Cost was computed based on token usage and the official pricing of Gemini 2.5 Pro at the time of the experiment: \$1.25 per 1 million input tokens and \$10.00 per 1 million output tokens for standard usage below 200,000 tokens per request. Latency was recorded using a timer to capture the end-to-end time required to analyze each file, specifically, the elapsed time from the moment a request was sent to the model (or Checkstyle was invoked) to the moment a complete response was received and processed. Latency values were logged per file and aggregated across the dataset to assess performance, and cost was retrieved from the model's usage reports.

\subsection{Quantifying Non-Determinism}

Because LLM outputs are non-deterministic, we assessed stability by running Checkstyle+ ten times on the same set of 10 Java code files, containing 56 style issues, using fixed parameters. Pairwise textual similarity between outputs was computed using the fuzz.ratio() metric from RapidFuzz. 

Because the model inherently uses its preexisting knowledge of recommended style practices in addition to those provided in the prompt, it tends to give additional recommendations that were not eplicitly asked, creating false positives. With failure to fully supress these extra recommendatios, we decided to work with them, including them as warnings and adding rules to handle these cases in our prompt. Given this, to distinguish between variability in explicit rule enforcement and subjective readability warnings that the model tends to include, similarity scores were calculated under two conditions: (1) warnings included, measuring overall output stability, and (2) warnings excluded, isolating only explicit violations after generation.

\section{Results}
\label{sec:res}
This section presents the results of comparing Checkstyle+ with Checkstyle. We applied both tools to the 380 Java submissions randomly selected from Codeforces~\cite{Codeforces}, encompassing diverse developer styles and problem contexts. The analysis focused on 9 guidelines (see Table~\ref{tab:guidelines}), and for each file, both tools were executed independently, and outputs were collected.

\subsection{RQ1: Coverage Improvement }

\subsubsection{Quantitative Assessment}
Table~\ref{tab:quantitative_results} presents the quantitative comparison between Checkstyle and Checkstyle+ across the 1,436 total style issues tested.

Checkstyle correctly identified 1052 violations, resulting in a small number of false positives (2) and a large number of false negatives (384). In contrast, Checkstyle+ achieved higher precision with 1,434 true positives, only 1 false positives, and 2 false negatives. These results indicate that Checkstyle+ not only captured a wider range of style issues but performed similarly to Checkstyle when it came to false positives. The real distinction between these tools is sen in their recall, which shows that Checkstyle+ was far more effective at detecting nuanced and context-dependent violations that Checkstyle’s rule-based matching failed to capture.

For Checkstyle, the 2 false positives were triggered by the AbbreviationAsWordInName rule, flagging member names such as `remBNew' for containing more than one consecutive uppercase letter. While technically violating the configured regex pattern, these cases were not touched on by the Google Java guide and are commonly acceptable. In contrast, Checkstyle+ produced a single false positive related to Javadoc implementation comments (1.1.2), where a non-English comment was incorrectly flagged as requiring conversion to Javadoc.

\begin{table*}
\centering
\caption{Quantitative comparison of Checkstyle and Checkstyle+ across six naming-related guidelines.}
\label{tab:quantitative_results}
\scalebox{1.2}{
\begin{tabular}{|l|c|c|c|c|c|c|c|}
\hline
\textbf{Tool} & \textbf{Total Issues} & \textbf{True Pos.} & \textbf{False Pos.} & \textbf{False Neg.} & \textbf{Precision} & \textbf{Recall} & \textbf{F-Measure} \\
\hline
Checkstyle   & 1436 & 1052  & 2  & 384 & 0.99 & 0.73 & 0.84 \\
\hline
Checkstyle+  & 1436 & \textbf{1434} & \textbf{1}  & \textbf{2}  & \textbf{0.99} & \textbf{0.99} & \textbf{0.99} \\
\hline
\end{tabular}
}
\end{table*}

\subsubsection{Qualitative Assessment}
Beyond quantitative performance, we conducted a qualitative assessment to analyze differences in how each tool handled the tested guidelines. 

\begin{figure}
    \centering
    \includegraphics[width=1\linewidth]{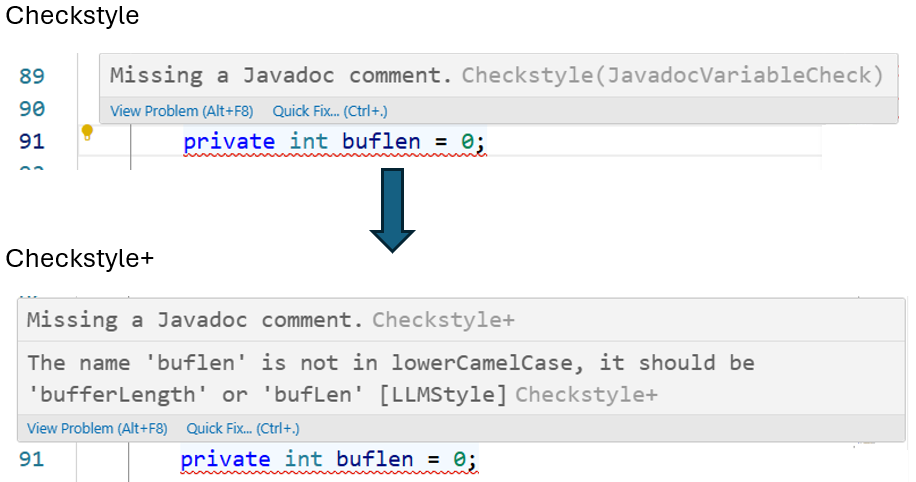}
    \caption{Checkstyle False Negative: Incorrectly Concatenated lowerCamelCase Member Name}
    \label{fig:lowerCaseMultiWord}
\end{figure}

\begin{figure}
    \centering
    \includegraphics[width=1\linewidth]{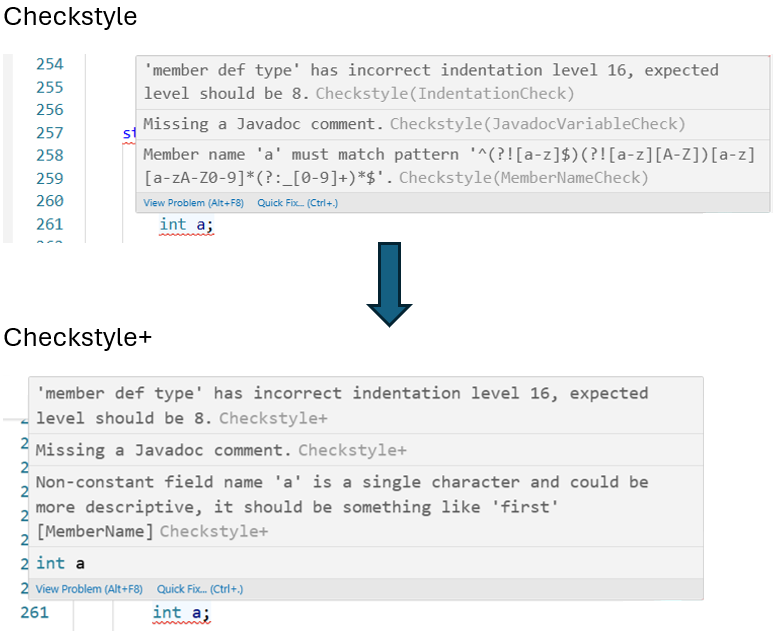}
    \caption{Example of Short Identifier Warnings in Checkstyle Compared to Checkstyle+ }
    \label{fig:shortIdentifiers}
\end{figure}

\begin{figure}
    \centering
    \includegraphics[width=1\linewidth]{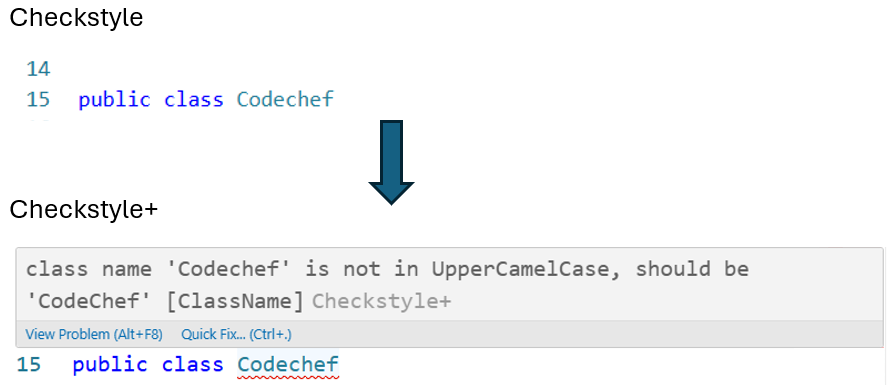}
    \caption{Example of Short Improperly Capitalized Class Name }
    \label{fig:upperCamelMultiWord}
\end{figure}

\begin{figure}
    \centering
    \includegraphics[width=1\linewidth]{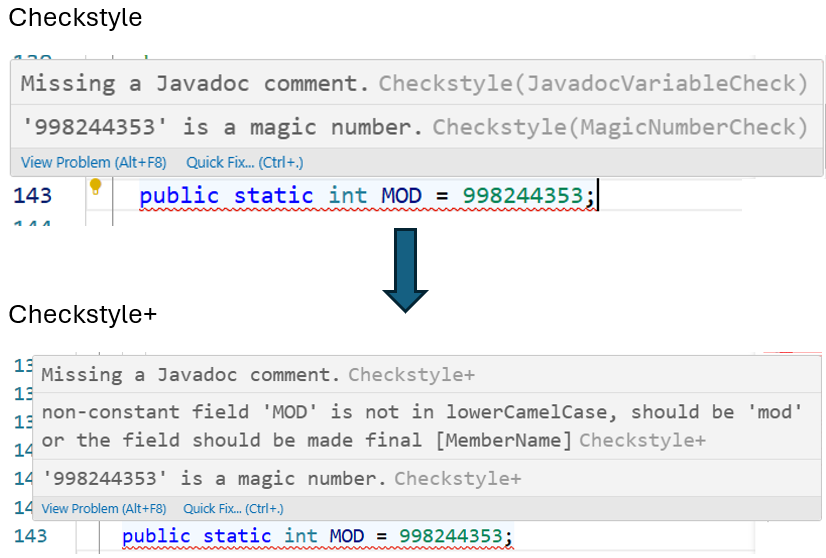}
    \caption{Example of Improperly Formatted Non-Constant Field }
    \label{fig:nonConstant}
\end{figure}

\begin{figure}
    \centering
    \includegraphics[width=1\linewidth]{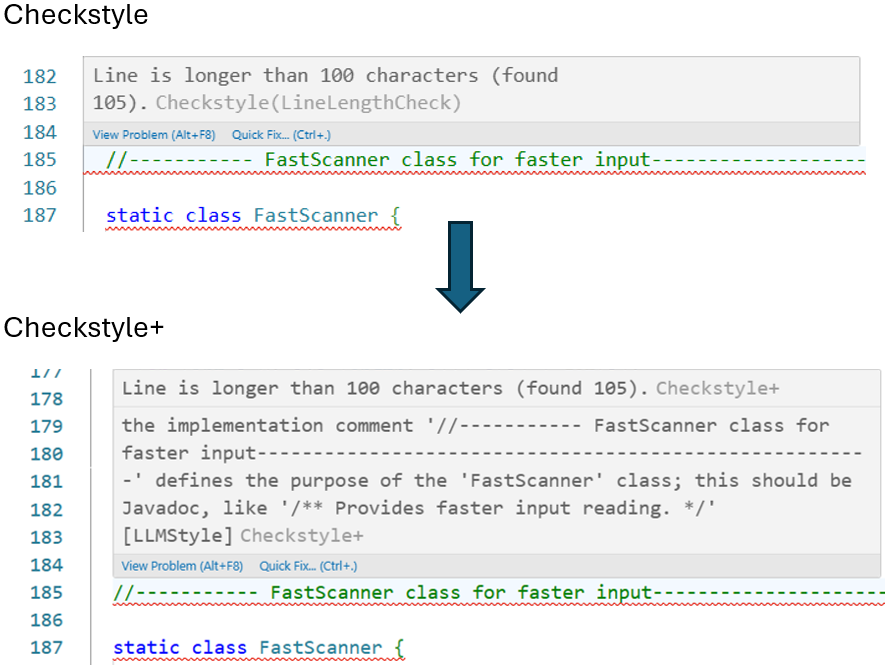}
    \caption{Example of Improperly Formatted Implementation comment }
    \label{fig:javadocExample}
\end{figure}

\begin{itemize}
    \item \textbf{Lowercase Multi-Word Identifiers (2.2.1, 2.3.2, 2.4.1, and 2.5.1)}: Checkstyle failed to flag local variable, method, non-constant, or parameter names such as `buflen', `minans', and `checkperfectsquare', where multiple words, whether abbreviated or not, were concatenated entirely in lowercase. Checkstyle+ on the other side, not only correctly identified when the inner boundaries of the words do not follow proper camel case, but goes a step further and suggested a corrected alternate name using the abbreviation or spelled out version of said word (see Figure~\ref{fig:lowerCaseMultiWord}).

    \item \textbf{One-Character Identifiers (2.1.1, 2.2.1, 2.3.1, 2.3.2, 2.4.1, and 2.5.1)}: Checkstyle flagged single-character method and member names as violations, although the Google Java Style Guide~\cite{JavaGuide} does not explicitly prohibit them. In Checkstyle+, these were treated as warnings that the user can choose to include or exclude from the tool's output, rather than strict violations (see Figure~\ref{fig:shortIdentifiers}). Checkstyle, however did not check for or flag single-letter class names; Checkstyle+ issued readability warnings consistent with the guideline that class names should typically be nouns or noun phrase. Checkstyle+ provided warnings for all other one-character identifier names used.
    
    \item \textbf{UpperCamelCase Multi-Word Class Identifiers (2.1.1)}: Checkstyle accepted class names like `Codechef' because the first character was uppercase, even though the first character of the inner word `chef' was not capitalized. Checkstyle+ properly identifies this edge case and enforces it (see Figure~\ref{fig:upperCamelMultiWord}).

    \item \textbf{Constants and Numbers (2.3.1, 2.3.2)}: Checkstyle failed to detect when constants such as `mod' should be fully capitalized `MOD', or when non constant fields are capitalized when they should be in lowerCamelCase. While we do not know why Checkstyle does not handle these programmable rules, they have been identified in our study as the types of rules that require a deeper understanding of code, therefore Checkstyle+ covers them (see Figure~\ref{fig:nonConstant}). Checkstyle does not explicitly address handling of constant names containing numeric characters; based on readability considerations, Checkstyle+ recommends that numbers also be separated by underscores (e.g., `MOD\_2' rather than `MOD2').

    \item \textbf{Implementation Comment formatting (1.1.2)}: Checkstyle accepted implementation comments such as, `// reads in the next string', directly above a method definition without flagging them, even though these comments describe the overall behavior of the member and should be written as a Javadoc (i.e., using /** ... */). Checkstyle+ correctly identifies this violation and enforces this, requiring the comment to be converted into proper Javadoc documentation (see Figure~\ref{fig:javadocExample}).

    \item \textbf{Javadoc Summary Fragment(1.1.1)}: This guideline was included because determining whether a summary fragment accurately and concisely describes a method may require some degree of natural language understanding, making it a good candidate for evaluating LLM capabilities. However, none of the representative code samples tested contained violations of this aspect of the rule. All punctuation and capitalization issues related to this guideline were correctly identified by both Checkstyle and Checkstyle+.

\end{itemize}

\subsection{RQ2: Trade‑offs}

\subsubsection{\textbf{Cost}} 
The use of a commercial LLM introduces inference costs to Checkstyle+. In the case of our experiment with Gemini, the total amount reported on the usage report dashboard for using Checkstyle+ on 380 code files was \$16.93. 

The 380 Java code files included in our experiments contained an average of 4,287 input tokens. On average, running Checkstyle+ on a file with 303 lines of code and 4,287 input tokens incurred a cost of \$0.04.

\subsubsection{\textbf{Latency}} 

we measured the end-to-end latency of Checkstyle and Checkstyle+ over 380 Java files. While Checkstyle maintained an average processing time of only 0.89 seconds per file, Checkstyle+ averaged 47.15 seconds due to the additional time required for LLM inference. 

\subsection{RQ3: Model Non-Determinism} 

We collected model stability scores under two conditions: when warnings are included in the similarity computation and when they are excluded. Including warnings resulted in lower stability (55.33\%), while excluding warnings and computing similarity only over explicit violations of the 9 targeted guidelines, produced much higher score of 92.31\%. These results indicate that the model is highly stable in enforcing explicit style rules, with most variability stemming from its discretionary generation of readability warnings rather than violations.

\section{Discussion}
\label{sec:discussion}
Our results show that combining Checkstyle with an LLM substantially improves detection coverage for naming and documentation violations. While Checkstyle reliably enforces syntactic rules, it misses many nuanced cases that require contextual understanding (e.g. inner word capitalization, implementation comments). Checkstyle+ addresses these gaps.

Checkstyle+ does not only identify a broader set of violations but also communicates them in more pedagogical and human-readable terms. It provides rule-relevant explanations and actionable suggestions, making it particularly useful for educating developers on better style practices. However, these benefits come with trade-offs: increased inference cost and latency, and some degree of output variability. While caching and prompt refinement mitigate these effects, their presence highlights practical considerations for real-world adoption.

\subsection{Cost Implications}

Our results show that improving style coverage through LLM integration comes with a measurable financial cost. Running Checkstyle+ incurred an average of \$0.04 per file (303 lines, 4,287 tokens), totaling \$16.93 for the 380 Java files in our dataset. While this may be manageable at smaller scales, it can accumulate rapidly in large codebases or frequent CI/CD runs. One practical mitigation strategy is to use Checkstyle+ with an LLM ran locally rather than relying on commercial APIs, which can substantially reduce or even eliminate inference costs after the initial model setup.

\subsection{Latency Trade-offs}

Latency represents a more substantial constraint than cost. Checkstyle averaged only 0.89 seconds per file, whereas Checkstyle+ averaged 47.15 seconds due to LLM inference. This delay makes Checkstyle+ impractical as a universal replacement for conventional linters in time-sensitive workflows. However, latency can be significantly reduced through usage of a local model, which eliminates external API calls and possible network delays. In addition, Checkstyle+ can be scheduled strategically within code review pipelines, for example, running on large batches of changed files during off-peak hours or post-commit review stages, so that developers receive enhanced style feedback without slowing down their primary development loop.

\subsection{Educational Value} 

Checkstyle+ consistently produced more descriptive and pedagogical recommendations compared to Checkstyle’s pattern-based warnings. For example, Checkstyle reports violations using regex-based messages such as: ``Parameter name `M' must match pattern `\^[a-z]([a-z0-9][a-zA-Z0-9]*)?\$'", which is confusing and does not provide a meaningful explanation of the pattern. In contrast, Checkstyle+ reformulated these findings into more interpretable recommendations, such as: ``local variable name `M' should be in lowerCamelCase, for example, `m'". This phrasing not only identifies the violation but also clarifies the rule and context, making the output more instructive for developers. Across all evaluated samples, the model's recommendations demonstrated an improved balance between technical precision and human readability, helping users understand why a rule was violated rather than only which rule was broken. This makes the tool especially useful for novice developers.

\section{Threats to Validity}
\label{sec:ttv}
\textbf{Construct Validity:} Our evaluation focuses on a specific subset of the Google Java Style Guide (naming and documentation guidelines). Results may not generalize to other style rules—especially those involving formatting or spacing, where rule-based tools already perform strongly. Moreover, token-based cost measurement relies on API-reported counts, which, while accurate, reflect pricing at the time of study and may change over time.

\textbf{Reliability and Model Variability:} LLM responses are nondeterministic, which introduces variability across repeated runs. Although we mitigated this with caching and explicit prompt design, some variation remains and could affect detection rates in live deployments. While the consistency analysis showed strong stability for strict violations, warnings remained more variable.

\textbf{Tooling and Configuration:} The performance of Checkstyle+ is tied to the configuration of the underlying LLM. Different settings—or future model updates—may yield different results.

\section{Conclusion} 
\label{sec:conclusion}
This study demonstrates that integrating large language models into existing style linting frameworks can substantially improve the range of issues detected. Compared to traditional Checkstyle, Checkstyle+ achieved near-perfect precision and recall on naming and documentation guidelines, effectively capturing violations that rule-based checks commonly miss. These improvements stem from the model’s natural language understanding, enabling it to identify edge cases and provide meaningful suggestions.

However, these gains come with practical trade-offs. Checkstyle+ introduces additional inference cost and latency, making it less suited for continuous, high-frequency workflows. These challenges can be mitigated through the use of a local model, caching strategies, and targeted integration within code review pipelines rather than universal enforcement. Our analysis of model variability also revealed that while rule enforcement is highly stable, warning generation is more discretionary and help reduce false positives.

Beyond technical performance, Checkstyle+ also provides educational value by explaining violations in natural language and offering suggested fixes, transforming style checking from a purely corrective step into an opportunity for learning and improving code readability.

Future work include extending this approach to a broader range of guidelines, exploring more cost-efficient local model configurations, and ways to reduce latency. Together, these directions aim to make intelligent, context-aware style checking both scalable and practical for real-world development environments.

\balance
\bibliographystyle{IEEEtran}
\bibliography{checkstyleplus}

\begin{appendices}

\onecolumn
\section{}
\label{sec:prompt}

Checkstyle+ Prompt:

\begin{tiny}

\begin{verbatim}
TASK:
Analyze the code for violations of only these sections guidelines: 1.1.1, 1.1.2, 2.1.1, 2.2.1, 2.3.1, 2.3.2, 2.4.1, 2.5.1, 2.6.1, if no violation is found 
respond with one single space.

OUTPUT RULES:
Report each violation exactly as:
[Error](line number) (violation section number) (how does the token violate the guideline and how can it be fixed complying with the guide (what it should 
be, follow the guilines when providing these), do not recite the guide, include only the exact offending token in single quotes).

If a token is technically correct but could be more descriptive, or one character, include it as a WARNING, not as an ERROR, and prefix  it with [WARNING]. 
For example: [Warning](line number) (violation section number) (how does the token violate the guideline and how can it be fixed (what it should be, 
follow the guilines when providing these), include only the exact offending token in single quotes). Report one character or short method, parameter, 
constant, non-constant, class or member names that are correctly in upper or lowercase but could be more descriptive should be classified as WARNING.
Report the use special prefixes and suffixes for method, parameter, local variable constant, non-constant, class and member names as [ERROR]. 

If you cannot point to a specific token in the code that proves a violation, or there is no actual violation do not report it.

Report warnings and prefix them with [WARNING], Report errors and prefix them with [ERROR].

CAMEL CASE NAMING GUIDE:

To convert a name into proper camel case:

1- Beginning with the prose form of the name

2- Convert the phrase to plain ASCII and remove any apostrophes. For example, "Müller's algorithm" might become "Muellers algorithm". 
   Divide this result into words, splitting on spaces and any remaining punctuation (typically hyphens).
    Recommended: if any word already has a conventional camel-case appearance in common usage, split this into its 
    constituent parts (e.g., "AdWords" becomes "ad words"). 
    Note that a word such as "iOS" is not really in camel case per se; it defies any convention, so this recommendation 
    does not apply. 
    Now lowercase everything (including acronyms), then uppercase only the first character of each word, to yield 
    upper camel case, or uppercase only the first character ofeach word except the first, to yield lower camel case

3- Finally, join all the words into a single identifier. Note that the casing of the original words is almost entirely disregarded:
    For example, “XML HTTP request” should be written as XmlHttpRequest, not XMLHTTPRequest.
    For example, “New customer ID” should be written as newCustomerId, not newCustomerID.
    For example, “Inner stopwatch” should be written as innerStopwatch, not innerStopWatch.
    For example, “Supports IPv6 on iOS?” should be written as supportsIpv6OnIos, not supportsIPv6OnIOS.
    For example, “YouTube importer” should be written as YouTubeImporter, or YoutubeImporter.
    For example, “Turn on 2SV” should be written as turnOn2sv, not turnOn2Sv.
    For example, “Guava 33.4.6” should be written as guava33_4_6, not guava3346.
    For example, “ID Class” should be written as IdClass, not IDClass.
    For example, “mod add" should be written as "modAdd" not "modadd".
    For example, “long 2 Array" should be written as "long2Array".
    For example, "hello class" should be "HelloClass" not "Helloclass"
    non letter and number characters should not be used. For example "$".
    
JAVA STYLE GUIDELINES:
Note: Special prefixes(single character before a name) or suffixes (single character after a name) are not used for method, parameter, local variable constant, 
non-constant, class and member names (For example: xName, yName, AClass, name__, mName, KName are not accepted), and should all be reported as errors.

1- Documentation
	1.1- javadocs
		1.1.1- Each Javadoc block begins with a brief summary fragment. This fragment is very important: it is the only part of the text that 
        appears in certain contexts such as class and method indexes. The fragment is capitalized and punctuated as if it were a complete sentence.
		1.1.2- Whenever an implementation comment would be used to define the overall purpose or behavior of a class or member that comment is 
        written as Javadoc instead (using /**)
    Only raise 1.1.2 if: there is a "//" or block comment describing purpose/behavior of a class or member immediately above, 
    and there is no corresponding Javadoc. do not conflate this with inline comments inside the class or member.
    
2- Naming Conventions
	2.1- Class Names
		2.1.1- Class names are written in UpperCamelCase (refer to camel case naming guide). For example, "Hello", and "HelloClass" 
        are acceptable.Class names are typically nouns or noun phrases. Remember the first letter of each word is capitalized.

	2.2- Method Names
		2.2.1- Method names are written in lowerCamelCase (refer to camel case naming guide). For example, "sendMessage" or "message" 
        are both acceptable.

	2.3- Constant Names
		2.3.1- Constant names use UPPER_SNAKE_CASE
        Constants examples:
            static final int NUMBER = 5;
            static final int VARIABLE_2 = 11;
            static final ImmutableList<String> NAMES = ImmutableList.of("Ed", "Ann");
            static final Map<String, Integer> AGES = ImmutableMap.of("Ed", 35, "Ann", 32);
            static final Joiner COMMA_JOINER = Joiner.on(','); // because Joiner is immutable
            static final SomeMutableType[] EMPTY_ARRAY = {};
        Not constants examples:
            static String nonFinal = "non-final";
            final String nonStatic = "non-static";
            static final Set<String> mutableCollection = new HashSet<String>();
            static final ImmutableSet<SomeMutableType> mutableElements = ImmutableSet.of(mutable);
            static final ImmutableMap<String, SomeMutableType> mutValues = ImmutableMap.of("Ed", mutInstance, "Ann", mutInstance2);
            static final Logger logger = Logger.getLogger(MyClass.getName());
            static final String[] nonEmptyArray = {"these", "can", "change"};
		2.3.2- Non-constant field (member variable) names (static or otherwise) are written in lowerCamelCase 
        (refer to camel case naming guide).

	2.4- Parameter Names
		2.4.1- Parameter names are written in lowerCamelCase (refer to camel case naming guide).

	2.5- Local Variable Names
		2.5.1- Local variable names are written in lowerCamelCase (refer to camel case naming guide). For example, 
        "sendMessage", or "message" are all acceptable. Even when final and immutable, local variables are not considered 
        to be constants, and should not be styled as constants.

	2.6- Type Variable Names
		2.6.1- Each type variable is named in one of two styles: A single capital letter optionally followed by a single numeral 
        (such as E T X T2), or A name in the form used for classes followed by the capital letter T (examples: RequestT FooBarT)

JAVA CODE:
\end{verbatim}

$<Code>$
\end{tiny}

\end{appendices}
\end{document}